\documentstyle[aps]{revtex}
\begin{document}                
\def\be{\begin{equation}}
\def\ee{\end{equation}}
\def\bea{\begin{eqnarray}}
\def\eea{\end{eqnarray}}
\title{QUANTUM MECHANICAL CARRIER OF THE IMPRINTS OF GRAVITATION
\footnote{Published in Phys. Rev. D {\bf 57} 4718-4723 (1998) and based
on part of a report given in the session ``Quantum Fields in Curved Space
I'' originally published in the Proc. of the Eighth Marcel Grossmann
Meeting on General Relativity, Jerusalem, June 23-27, 1997, edited by
Tsvi Piran (World Scientific, Singapore, 1999), 806-808}
}

\author{ULRICH H. GERLACH}
\address{Department of Mathematics, Ohio State University, Columbus, OH
43210, USA}
%
\maketitle
\begin{abstract}                
We exhibit a purely quantum mechanical carrier of the imprints of
gravitation by identifying for a relativistic system a property which
(i) is independent of its mass and (ii) expresses the Poincare
invariance of spacetime in the absence of gravitation. This carrier
consists of the phase and amplitude correlations of waves in
oppositely accelerating frames. These correlations are expressed as a
Klein-Gordon-equation-determined vector field whose components are the
``Planckian power'' and the ``r.m.s. thermal fluctuation'' spectra.
The imprints themselves are deviations away from this vector field.
\end{abstract}

\begin{abstract}
\end{abstract}
\section{INTRODUCTION}

Does there exist a purely quantum mechanical carrier of the imprints
of gravitation? The motivation for considering this question arises
from the following historical scenario: Suppose a time traveller from
today visited Einstein in 1907, when he had the ``happiest thought of
his life'', just in time before Einstein actually started on the path
which led towards his formulation of gravitation (general relativity).
Suppose that this time traveller told Einstein about relativistic
quantum mechanics, about event horizons, and about Rindler's exact
spacetime coordinatization of paired accelerated frames.  Would that
visitor from today have been able to influence the final form of the subsequent
theory of gravitation?  Put differently, how different would the
course of history have been if Einstein had grafted relativistic
quantum mechanics and Rindler spacetime onto the {\it roots} of
gravitation instead of its trunk or branches?

\section{THE EOTVOS PROPERTY}

Relativistic quantum mechanics is a broad foundation of
physics, but on the surface it seems to have no bearing on
gravitation. Nevertheless, we shall choose this foundation as the
vantage point for viewing gravitation. In order not to appear
arbitrary, it is necessary that we remind ourselves of the key ideas
that led Einstein to his formulation of gravitation and then assess
how well they comply with relativistic quantum mechanics.

\subsection{Classical}

Gravitation is characterized by the fact that every test particle traces out
a world line which is independent of the parameters characterizing the
particle, most notably its mass. Thus, cutting the mass of
a planet into half has no effect on its motion in a gravitational
field. No intrinsic mass parameter is needed to specify a particle's
motion.  The world line is determined entirely by the particle's local
environment, not by the particle's intrinsic structure such as its
mass. We shall refer to this structure independence as the
``Eotvos property''
\cite{1}.

This property is the key to making gravitation subject to our
comprehension. This independence is far reaching, because, with it,
the world lines have a remarkable property: They are probes that
reveal the nature of gravitation without themselves getting affected
by the idiosyncracies (e.g. mass) of the particles. The
imprints of gravitation are acquired without the introduction of
irrelevant features such as the mass of the system that carries these
imprints.  Put differently, the classical world lines of particles
highlight a basic principle: Gravitation is to be identified by
its essentials.

Einstein used this principle to acquire our understanding about
gravitation within the framework of classical mechanics. His line of
reasoning was as follows:

(1) He identified an ``accelerated frame'' as a one-parameter family
of locally inertial (=free-float) frames
\cite{2}. 
The free-float nature of
each of these frames he ascertained strictly within the purview of
classical mechanics, namely by means of the straightness of all
particle world lines (Newton's first law of motion)
\cite{refTandW}.

(2) Next he observed that the afore-mentioned Eotvos property of
the particle world lines implies and is implied by the statement
(equivalence principle) that the motion of particles, falling in what
was thought to be a non-accelerated frame with gravity present, is
physically equivalent to the motion of free particles viewed relative
to an accelerated frame 
\cite{4,5,6}.

(3) With this observation as his staring point, he used Lagrangian
mechanics to characterize gravitation in terms of the metric tensor
\cite{4,5,6}, 
and then proceeded towards his theory of gravitation
\cite{7} 
along what in hindsight is a straight forward mathematical
path 
\cite{8}

However happy we must be about Einstein's gravitation theory, we must
not forget that it rests on two approximations, which, although very
fruitful, are approximations nevertheless. They are made in the very
initial stages of Einstein's line of reasoning, namely in {\it what}
he considered an ``accelerated frame'' and in {\it how} he used it.

First of all, in step (1) above, Einstein approximates an
``accelerated frame'' as one in which its future and past event
(Cauchy) horizons are to be ignored.  Such indifference is
non-trivial.  It results in neglecting the fact that (i) {\it a frame
accelerating uniformly and linearly always has a} twin, {\it which
moves into the opposite direction} and (ii) that it takes these {\it two}
frames (``Rindler frames'') to accomodate a Cauchy hypersurface.

Secondly, the geodesic world lines, the carriers of the imprints from
which he constructs his theory of gravitation, are only classical
approximations to the quantum mechanical Klein-Gordon wave
functions. Their domain extends over all (the regions) of spacetime
associated with the pair of accelerated frames. The world lines, by
contrast, have domains which are strictly limited to the razor sharp
classical particle histories.

It would have been difficult to argue with these
approximations ninety years ago. However, the knowledge we have gained in the
meantime would expose us to intellectual evasion if we were to insist
on adhering to them in this day and age. Thus we shall not make them.

\subsection{Quantum Mechanical}

Even though we shall distance ourselves completely from these two
approximations, we shall adhere to the above-mentioned principle in
order to make gravitation comprehensible. This means that we shall
bring the Eotvos property into the purview of quantum mechanics by
insisting that we find a carrier of the imprints of gravitation which
is {\it both} independent of the intrinsic mass of the carrier, {\it and}
is purely quantum mechanical in nature. 

This requirement turns out to be extremely restrictive, but not overly
so.

In quantum mechanics, unlike in classical mechanics, the dynamics of a
system is an issue separate from the measurement of its
properties. Consequently, our formulation comes in two parts. There is
the mathematical part which develops the dynamics of the carrier, and
there is the physical part which describes how to measure the
gravitational imprints with physically realizable apparatus.  In this
article we shall consider primarily the first part. A key aspect of the 
second is pointed out in the concluding section, while a more extensive 
sketch has been consigned to another article.
\cite{9}.

\section{ACCELERATION-INDUCED CARRIER OF THE IMPRINTS OF GRAVITATION}

The most obvious candidate for such a carrier is the set of wave
functions of a particle. They take over the role that the particle
world lines occupy in classical mechanics. Indeed, in the classical
limit, any world line can be recovered from the wave functions via the
principle of constructive interference.  We shall find that the
imposition of the Eotvos property on the wave functions is not
only extremely restrictive, but also forces us into viewing spacetime
from a perspective which is very different from the customary one.

First recall the de Broglie relations. They relate the wavelength and
the frequency of a particle's wave function to its momentum and
energy, and to its mass. Next recall that, quite generally, the set of
dynamical constants of motion of particles not only characterizes the
particles, but also indicates the nature of the reference frame within
which they move. In fact, there is a one-to-one correspondence between
the constants of motion (``complete set of commuting observables'')
and the type of reference frame in which they are observed. Thus
conservation of momentum and energy for all particles implies that
they are observed in a free-float (inertial) reference frame.
Consequently, if one chooses to describe the quantum dynamics of the
particles in the momentum representation, then, by default, one has
chosen to observe these particles relative to a free-float frame.
This means that the choice of the representation characterized by a
complete set of commuting observables (here the momentum and the
energy) goes hand in hand with the customary view of observing
spacetime from the perspective of free float frames.

In the momentum representation the wave function depends explicitly
on the mass of the particle. Consequently, this wave function does not
qualify as a carrier of the imprints of gravitation.

We shall now remedy this deficiency by exhibiting the
quantum dynamics of the particle in a new representation relative to
which the wave function does have the Eotvos property.  The
spacetime perspective of this representation consists of the familiar
four Rindler sectors induced by two (noninertial) frames accelerating
into opposite directions.

\subsection{Relativistic Quantum Mechanics}
Our starting point is non-trivial relativistic quantum mechanics as
expressed by the Klein-Gordon equation
\begin{equation}
{\partial^2 \psi \over \partial t^2}-
{\partial^2 \psi \over \partial z^2}+ k^2\psi =0,
\label{eq:1}
\end{equation}
where $k^2=k^2_x +k^2_y +m^2 $.

The objective is to deduce from this equation a
carrier of the imprints of gravitation with the following three
fundamental requirements:

1. The imprints must be carried by the evolving dynamics of a quantum
mechanical wavefunction.

2. Even though the dynamical system is characterized by its mass
$m$, the carrier and the imprints must {\it not} depend on this mass, i.e. 
the carrier
must be {\it independent} of $k^2$. This requirement is analogous to the
classical one in which the world line of a particle is independent of
its mass.

3. In the absence of gravitation the carrier should yield measurable
results (expectation values) which are invariant under Lorentz boosts
and spacetime translations.

We shall now expand on these three requirements.
In quantum mechanics the wave function plays the role which in
Newtonian mechanics is played by a particle trajectory or in
relativistic mechanics by a particle world line. That the wave
function should also assume the task of carrying the imprints of
gravitation is, therefore, a reasonable requirement.

Because of the Braginski-Dicke-Eotvos experiment, the motion of bodies
in a gravitational field is independent of the composition of these
bodies, in particular their mass. Consequently, the motion of free
particles in spacetime traces out particle histories whose details
depend only on the gravitational environment of these particles, not
on their internal constitution (uniqueness of free fall, ``weak
equivalence principle''). Recall the superposition of different wave
functions (states) of a relativistic particle yields interference
fringes which do depend on the mass of a particle (``incompatibility
between quantum and equivalence principle''
\cite{10}). 
If the task of
these wave functions is to serve as carriers of the imprints of
gravitation, then, unlike in classical mechanics, these interfering
wave functions would do a poor job at their task: They would respond
to the presence (or absence) of gravitation in a way which depends on
the details of the internal composition (mass) of a particle. This
would violate the simplicity implied by the Braginski-Dicke-Eotvos
experiment. Thus we shall not consider such carriers. This eliminates
any quantum mechanical framework based on energy and momentum
eigenfunctions because the dispersion relation, $E^2=m^2 +p^2_z +p^2_y
+p^2_x$, of these waves depends on the internal mass $m$.

Recall that momentum and energy are constants of motion which imply
the existence of a locally inertial reference frame. Consequently,
requirement 2. rules out {\it inertial} frames as a viable spacetime
framework to accomodate any quantum mechanical carrier of the imprints
of gravitation.  Requirement 2. also rules out a proposal to use the
interference fringes of the gravitational Bohm-Aharanov effect to
carry the imprints of gravitation
\cite{11}. This is because the fringe
spacing depends on the rest mass of the quantum mechanical particle.

Requirement 3. expresses the fact that the quantum mechanical carrier
must remain unchanged under the symmetry transformations which characterize
a two-dimensinal spacetime. By overtly suppressing the remaining two spatial
dimensions we are ignoring the requisite rotational symmetry. Steps towards
remedying this neglect have been taken elsewhere
\cite{12}.

We shall now exhibit a carrier which fulfills the three fundamental
requirements. It resides in the space of Klein-Gordon solutions whose
spacetime domain is that of a {\it pair} of frames accelerating into
opposite directions (``Rindler frames'').  These frames partition
spacetime into a pair of isometric and achronally related Rindler
Sectors $I$ and $II$,
\begin{equation}
\left. 
\begin{array}{c}
  t-t_0 =\pm\xi\sinh \tau \\
  z-z_0  = \pm\xi  \cosh \tau
  \end{array} 
\right\} \quad 
  \begin{array}{l}
  +:~~\hbox{``Rindler Sector I''} \\
  -:~~\hbox{``Rindler Sector II''} \quad .
  \end{array}
\label{eq:coordinate transform}
\end{equation}
Suppose we represent an arbitrary solution to the K-G equation in the form
of a complex two-component vector normal mode expansion
\begin{equation}
\left(
\begin{array}{c}
\psi _I (\tau,\xi) \\
\psi_{II}(\tau,\xi)
\end{array}
\right)
= \int ^\infty _{-\infty} \{ a_\omega
\left(
\begin{array}{c}
1 \\
0
\end{array}
\right)
+b^*_\omega
\left(
\begin{array}{c}
0 \\
1
\end{array}
\right) \}\sqrt{2\vert \sinh \pi \omega \vert}
{K_{i\omega}(k\xi)\over\pi}e^{-i\omega\tau} d\omega 
\equiv \int ^\infty _{-\infty} \psi_\omega d\omega
\quad . \label{eq:modeintegral}
\end{equation}
This is a {\it correlated} (``entangled'') state with two independent
degrees of
freedom. There is the {\it polarization} degree of freedom in addition
to the spatial degree of freedom. The polarization degree of freedom
has a two-dimensional space of states ($C^2$) spanned by two-spinors.
The two components of a spinor
refer to the wave amplitude at diametrically opposite events on a
Cauchy hypersurface $\tau=constant$ in Rindler $I$ and $II$
respectively. The spatial ($0<\xi<\infty$) degree of freedom has a 
state space ($\cal H$) which is $\infty$-dimensional and which is 
spanned by the scalar boost eigenfunctions
\begin{equation}
\phi _\omega= \sqrt{2\vert \sinh \pi \omega \vert}~
{K_{i\omega}(k\xi)\over\pi}~e^{-i\omega\tau} \quad ,
\end{equation}
solutions to the Rindler wave equation
\be
\left[{1\over {\xi ^2}}{\partial^2\over \partial\tau^2}-{1\over
\xi}{\partial\over\partial\xi}\xi{\partial\over \partial\xi}
+k^2\right]
\phi_{\omega}(\tau ,\xi )=0 \quad ,
\ee
which is the equation obtained by applying the coordinate transformation
Eq.(\ref{eq:coordinate transform}) to the Klein-Gordon Eq.(\ref{eq:1}).

\subsection{Geometry of the Space of Solutions}

The representation (\ref{eq:modeintegral}) puts us at an important mathematical juncture:
We shall forego the usual picture of viewing this solution as an
element of Hilbert space ($C^2\otimes \cal H$) with the Klein-Gordon
inner product,
\begin{eqnarray}
\lefteqn{ 
          {i\over 2}\int^0_{\infty}
          \psi^{\ast}_{II}{\stackrel{\leftrightarrow}{\partial} \over
          \partial\tau}\psi_{II}{d\xi\over\xi}
          +{i\over2}\int^{\infty}_0
          \psi^{\ast}_{I}
          {\stackrel{\leftrightarrow}{\partial} \over\partial\tau}\psi_I 
          {d\xi\over\xi} }                          \nonumber \\
& & = \int^{\infty}_{-\infty}\int^{\infty}_{-\infty} (a^*_\omega b_\omega )
         \left[ 
               \begin{array}{cc}
               1 & 0 \\
               0 &-1 
               \end{array}
         \right] \left( 
                       \begin{array}{c}
                       a_{\omega'} \\
                       b^*_{\omega'}
                       \end{array}
                 \right) {i\over2}\int^{\infty}_0 \phi^{\ast}_{\omega}
   {\mathop{\partial}\limits^\leftrightarrow \over\partial\tau}\phi_{\omega'} 
   {d\xi\over\xi}                                  \nonumber \\
& &  = \int^{\infty}_{-\infty}\int^{\infty}_{-\infty}
       (a^*_\omega b_\omega )
                            \left[ 
                                  \begin{array}{cc}
                                  1& 0 \\
                                  0&-1
                                  \end{array}
          \right] \left( 
                        \begin{array}{c} 
                        a_{\omega'} \\
                        b^*_{\omega'}
                        \end{array}
                  \right)  {\omega \over \vert \omega \vert}
                           \delta (\omega -\omega')~ d\omega d\omega' ~
                     \begin{array}{c}
                     \hbox {(inner~product}\\ 
                     \hbox{ for}~C^2\otimes \cal H~)
                     \end{array}                   \nonumber \\
& & \equiv \int^{\infty}_{-\infty}\int^{\infty}_{-\infty}
            \langle\psi_\omega,\psi_{\omega '}\rangle~ d\omega d\omega'\quad .
\label{eq:innerproduct}
\end{eqnarray}
Instead, we shall adopt a qualitatively new and superior viewpoint. 
{\it Each Klein-Gordon solution is a spinor field over the Rindler 
frequency domain.} This is
based on the vector bundle $C^2\times R$.  Here $C^2$ is the complex
vector space of two-spinors, which is the fiber over the
one-dimensional base manifold $R=~\{ \omega:~-\infty < \omega < \infty
\} $, the real line of Rindler frequencies in the mode integral,
Eq.(\ref{eq:modeintegral}).

We know that one can add vectors in the {\it same} vector (fiber)
space.  However, one may not, in general, add vectors belonging to
different vector spaces at different $\omega$'s. The exception is when
vectors in different vector spaces are {\it parallel}. In that case
one may add these vectors.  The superposition of modes, Eq.(\ref{eq:modeintegral}),
demands that one do precisely this in order to obtain the two
respective total amplitudes of Eq.(\ref{eq:modeintegral}). In
brief, we are about to show that that \emph{the linear superposition principle
determines a unique law of parallel transport}.

The mode representation of Eq.(\ref{eq:modeintegral}) determines two parallel basis spinor
fields over $R$,
\begin{equation}
\lbrace \left( 
\begin{array}{c}
1 \\
0
\end{array}
\right)  \phi_\omega :~-\infty < \omega < \infty \rbrace
~~\hbox{and}~~
\lbrace \left( 
\begin{array}{c}
0 \\
1
\end{array}
\right)  \phi_\omega :~-\infty < \omega < \infty \rbrace \quad ,
\label{eq:basis}
\end{equation}
one corresponding to ``spin up'' ($\psi$ has zero support in Rindler $II$), 
the other to ``spin down'' ($\psi$ has zero support in Rindler $I$). 
This parallelism is dictated by the superposition principle, Eq.(3):
The total amplitude at a point $(\tau,\xi)$ in Rindler $I$ (resp. $II$)
is obtained by adding all the contributions from Rindler $I$ (resp. $II$)
only.

One can see from Eq.(\ref{eq:innerproduct}) that, relative to this
basis, the fiber metric over each $\omega$ is given by
\begin{equation}
(a^*_\omega b_\omega )\left[ 
\begin{array}{cc}
1&0 \\
0&-1 
\end{array}
\right] \left( 
\begin{array}{c}
a_\omega' \\
b^*_{\omega'}
\end{array}
\right)=a^*_\omega a_\omega -b_\omega b^*_\omega;~~~-\infty < \omega < \infty
\quad .
\label{eq:fibermetric} 
\end{equation}

It is evident that the law of parallel transport defined by
Eq.(\ref{eq:basis}) is {\it compatible} with this
(Klein-Gordon induced) fiber metric. This is because the two parallel
vector fields, which are represented by
\begin{equation}
\left( 
\begin{array}{c}
1 \\
0
\end{array}
\right) ~~\hbox{and}~~
\left( 
\begin{array}{c}
0 \\
1
\end{array}
\right);    ~-\infty < \omega < \infty 
\label{eq:parallelvectors}
\end{equation}
relative to the basis (\ref{eq:basis}), are orthonormal with respect to the 
the fiber
metric Eq.(\ref{eq:fibermetric}) at each point $\omega$ of the base space $R$.
It is not difficult to verify that these two spinor fields
are (Klein-Gordon) orthonormal in each fiber over $R$. The spinor
field
\begin{equation}
\{ \left(
\begin{array}{c}
a_\omega \\
b_\omega^*
\end{array}
\right) : ~-\infty < \omega < \infty \} \quad 
\end{equation}
is a section of the fiber bundle $C^2\times R$ and it represents a
linear combination of the two parallel vector fields. It is clear that
there is a {\it one-to-one correspondence between $\Gamma(C^2\times R)$,
the $\infty$-dimensional space of sections of this spinor bundle, and
the space of solutions to the Klein-Gordon equation.}

\subsection{Quantum Mechanical Carrier of Gravitational Imprints}

Our proposal is
to have each spinor field serve as a carrier of the imprints of
gravitation: A gravitational disturbance confined to, say, Rindler
$I$ or $II$ would leave its imprint on a spinor field at
$\tau=-\infty$ by changing it into another spinor field at
$\tau=+\infty$.

We know that in the absence of gravitation each of the positive and
negative Minkowski plane wave solutions evolves independently of all
the others.  This scenario does not change under Lorentz boosts and
spacetime translations.  This is another way of saying that the system
described by these solutions is Poincare invariant. Will the proposed
carriers comply with this invariance, which is stipulated by
fundamental requirement 3.?  To find out, consider a typical plane
wave. Its spinor representation (\ref{eq:modeintegral}) is
\be
\left[
\begin{array}{c}
e^{-i(t-t_0)k\cosh \theta +i(z-z_0)k\sinh \theta}\vert _I \\
e^{-i(t-t_0)k\cosh \theta +i(z-z_0)k\sinh \theta}\vert _{II} 
\end{array}
\right] =\int\limits^{\infty}_{-\infty}
e^{i\omega\theta}
\left[
\begin{array}{c}
e^{\pi\omega/2} \\
e^{-\pi\omega/2}
\end{array}
\right] {K_{i\omega}(k\xi )\over \pi} e^{-i\omega\tau} d\omega \quad .
\ee
This is a state with 100\% correlation between the boost energy and
the polarization (``spin'') degrees of freedom.  In recent years such
states have been called ``entangled'' states\cite{Peres}.
Suppose that for each
boost energy we determine the normalized Stokes parameters of this
polarization, i.e. the expectation values of the three modified Pauli
``spin'' matrices
\be
{\overrightarrow \sigma}:
     \{\sigma_1,\sigma_2,\sigma_3\}=\left\{\left[ 
\begin{array}{cc}
 0&1 \\
-1&0
\end{array}
\right],
\left[
\begin{array}{cc}
0&-i \\
-i&0
\end{array}
\right],
\left[
\begin{array}{cc}
1&0 \\ 
0&-1
\end{array}
\right]\right\} \quad .
\ee
This is a
three-dimensional vector field over the base manifold $R$, and is given by
\cite{12}
\be
       {{\langle\psi_\omega,{\overrightarrow \sigma}  \psi _{\omega '}
\rangle}\over {\langle\psi_\omega,\psi_{\omega '}\rangle}}
=\pm \left( \sqrt{N(N+1)},0,{1\over 2}+N \right);~~~
N=(e^{2\pi \omega} -1)^{-1};~~-\infty<\omega < \infty
\label{eq:imprintcarrier}
\ee
In compliance with requirements 2. and 3., this vector field is (a)
{\it independent of the particle mass} and (b) the same for {\it all}
positive (negative) Minkowski plane wave modes, a fact which expresses
its Poincare invariance. The presence of {\it gravitation would leave
its imprints by producing characteristic alterations in this vector
field.}

\section{SUMMARY}

\subsection{This Article}

Equation (\ref{eq:imprintcarrier}) is the center of a constellation
consisting of the following three results:
 
The main result is the recognition of the fact that a Klein-Gordon
charge has a property which displays the Poincare invariance of
Minkowski spacetime {\it without involving its specific mass}. This
property is the set of expectation values given by
Eq.(\ref{eq:imprintcarrier}). The presence of gravitation is expressed
by distortions of this mass-independent vector field. This property is
a quantum mechanical extension of the principle familiar from
classical mechanics, and it is dictated by the Braginski-Dicke-Eotvos
experiment, that the set of particle trajectories serve as the carrier
of the imprints of gravitation.

The second result is the fact that Eq.(\ref{eq:imprintcarrier}) are (twice) the
expectation values of the ``spin'' component operators
\be
{\bf L}:~L_1={\sigma _1\over 2},~L_2={\sigma _2\over 2},~
L_3={\sigma _3\over 2} \quad . 
\ee
Their commutation relations
\bea
[L_1,L_2]&=&-iL_3\cr
[L_2,L_3]&=&iL_1\cr
[L_3,L_1]&=&iL_2\quad ,
\eea
are those of the symmetry group $SU(1,1)$, which is precisely the
invariance group of the fiber metric, Eq.(\ref{eq:fibermetric}).  The
fact that $SU(1,1)$ is the invariance group of the single charge
system extends into the classical regime: If one interprets the
Klein-Gordon wave function as a classical field, then this $SU(1,1)$
symmetry gives rise 
\cite{12} 
to a {\it conserved vectorial ``spin''},
whose density is given by one half the expectation value of $\sigma$,
as in the numerator of Eq.(\ref{eq:imprintcarrier}).

The third result is as obvious as it is noteworthy: The
components of the vector field coincide with the ``Planckian power''
and the ``r.m.s.  thermal fluctuation'' spectra, in spite of the fact
that we have not made any thermodynamic assumptions. In fact, we are 
considering only the quantum mechanics of a single charge,
{\it or} the Klein-Gordon dynamics of a classical wave field. That
these two spectra also arise within the framework of a strictly
classical field theory, has already been observed in 
\cite{13}. 
We would
like to extend this observation by pointing out that in the absence of
gravitation these two spectra express the invariance of spacetime
under translations and Lorentz trasformations (requirement 3.,
Poincare invariance)

\subsection{The Wider Perspective}

We comprehend gravitation in two stages. First we identify the agent
which carries the imprints of gravitation. In Newton's formulation
this agent is the set of particle trajectories, in Einstein's
formulation the set of particle world lines, and in the quantum
formulation the set of correlations between the wave amplitudes in a
pair of oppositely accelerating Rindler frames.

In Newton's formulation the imprints consist of the bending of
the particle trajectories, in Einstein's formulation they consist of
the deviations of the geodesic world lines, and in the quantum formulation
they consist of the deviations of the correlations away from the 
``Planckian'' and the ``fluctuation'' spectral values given by
Eq.(\ref{eq:imprintcarrier}).

The second stage of our comprehension consists of relating these imprints
to the source of gravitation. In Newton's formulation this relation
is the Poisson equation, in Einstein's formulation the field equations of
general relativity, and in the quantum formulation we do not know
the answer as yet.

The equations for gravitation are an expression of the relation between the
properties inside a box and the resulting geometrical imprints of
gravitation on the surface of the box which surrounds the matter
source of gravitation. In the Newtonian theory the mass inside the box is
proportional to the total amount of gravitational force flux through the
surface of the box. In Einstein's theory the amount of energy and momentum
in the box is proportional to the total amount of moment of rotation on 
the surface of the box\cite{Wheeler,MTW}.

Alternatively, if the box is swept out by a coplanar collimated set of
moving null particles (e.g. neutrino test particles), then, upon
moving from one face of the box to the other, the neutrino pulse gets
focussed by an amount which is proportional to the amount of matter
inside the swept-out box. This proportionality, when combined with
energy-momentum conservation, is expressed by the Einstein field
equations\cite{Jacobson}. In this formulation the directed neutrino pulse is a
classical (i.e. non-quantum mechanical) carrier of the imprints of
gravitation, and the amount by which the pulse area gets focussed is
the gravitational imprint which is also classical. By letting the
neutrino pulse go into various directions, one obtains the various
components of the Einstein field equations.

It is interesting that the logical path from the classical imprints to
these field equations consists of a temporary excursion into the
quantum physics relative to an accelerated frame.  This excursion starts
with
the demand that one describe the state of the matter
inside the box relative to the frame of a uniformly accelerated
observer. The acceleration of this frame is to be collinear with the motion 
of the neutrino pulse. If
one complies with this demand, then the matter passing through the
neutrino pulse (event horizon) area is a flow of heat energy relative
to the accelerated frame. The temperature is the acceleration
temperature given by the Davies-Unruh formula. This permits one to
assign an entropy to the matter inside the box swept out by the
neutrino pulse. Consequently, the Rindler (boost) heat energy in the box is
the product of the matter entropy times the acceleration temperature.

Finally, the Einstein field equations follow from the 
Bekenstein hypothesis\cite{Bekenstein} that the entropy be proportinal to the
change of the area of the neutrino pulse area as it passes
through the box. 
 
It is obvious that this deduction of the field equations from the
Bekenstein hypothesis is only a temporary excursion into the
quantum physics relative to an accelerated frame. Indeed,
even though the proportionality between entropy and area consists of the 
squared Plack-Wheeler length, reference to Planck's constant $\hbar$
gets cancelled out by the Davies-Unruh temperature in the product 
which makes up  
the Rindler heat flux, the source for the Einstein field equations.
Consequently, the disappearance $\hbar$ from the heat flux guarantees 
that $\hbar$ will not appear in the Einstein field equations.

It seems evident that a quantum mechanical comprehension of gravitation
must start with a purely quantum mechanical carrier of its imprints.

\section{CONCLUDING REMARK}

Recall that in quantum mechanics, unlike in classical mechanics, the
problem of the dynamics of a system and the measurement of its
properties are two different issues. The dynamics is governed by a
differential equation, in our case the system's wave equation,
Eq.(\ref{eq:1}).  The measured properties are expressed by the
expectation values of the appropriately chosen operators.

It is obvious that our present treatment of these two
issues has been rather lopsided and we must address how to measure
the imprints of gravitation with physically realizable apparatus.
At first this seems like an impossibple task.

Consider the fact that that the quantum dynamics is governed by the
evolution of the Klein-Gordon wave function in the two Rindler frames,
Eq.(2), which (i) are accelerating {\it eternally} and (ii) are {\it
causally disjoint}.  How can one possibly find a physically realistic
observer who can access such frames? It seems one is asking for
the metaphysically impossible: an observer which can accelerate eternally
and can be in causally disjoint regions of spacetime. There simply is
no such observer!

However, we are asking the wrong questions because we have been
ignoring the two remaining Rindler sectors $P$ and $F$. 
\begin{equation}
\left. 
\begin{array}{c}
  t-t_0 =\pm\xi\cosh \tau \\
  z-z_0  = \pm\xi  \sinh \tau
  \end{array} 
\right\} \quad 
  \begin{array}{l}
  +:~~\hbox{``Rindler Sector F'' } \\
  -:~~\hbox{``Rindler Sector P''}
  \end{array}
\end{equation}
If one includes them into the identification of the carrier of
gravitational imprints, then our formulation leads to an astonishing
conclusion: the four Rindler sectors $I,II,P$ and $F$ together form a
{\it nature-given interferometer}, to be precise, the {\it Lorentzian}
version of a Mach-Zehnder interferometer
\cite{14}. 
The pair of oppositely 
accelerating frames $I$ and $II$ accomodate the pair of widely separated
coherent beams, and the two respective pseudo-gravitational potentials
serve as the two perfectly reflecting mirrors. The two Rindler sectors 
$P$ and $F$ serve as the respective half-silvered mirrors: $P$
acts as a beam splitter of radiation coming in from the past, while $F$ 
is where the two reflected beams interfere to produce a wave propagating into
the future.

Space limitations demand that we consign the task of elaborating on this
terse description to another article
\cite{9}. 
The point is that the quantum mechanical carrier, 
Eq.(\ref{eq:imprintcarrier}), has a firm physical foundation
while at the same time it exhibits the Eotvos property of being independent
of the intrinsic mass of the quantum system.



\end{document}